\let\latexaddtocontents\addtocontents
\documentclass[submitting, floatfix]{nst_arxiv}
\let\addtocontents\latexaddtocontents
\usepackage{subfigure,dcolumn}
\usepackage[T2A,T1]{fontenc}
\usepackage[english]{babel}
\usepackage{color}
\usepackage{listings}
\usepackage{natbib}
\let\footnote\relax

\let\textcite\relax

\let\citeauthor\relax
\let\citeyear\relax
\expandafter\let\csname
ver@natbib.sty\endcsname\relax

\usepackage[
    backend=biber,
    bibencoding=utf8,
    style=numeric-comp,
    giveninits=true,
    autolang=other, 
    sorting=none,
    doi=true, isbn=false, eprint=false, url=false, date=year, 
    maxcitenames=3,
    mincitenames=3,
    maxbibnames=3,
    minbibnames=3,
]{biblatex}

\usepackage{csquotes}
\renewbibmacro{in:}{}
\begin{document}
\renewcommand{\bibliography}[1]{}
\title{An event visualization software based on Phoenix for the CEPC experiment}
\author{Yu-Jie Zeng}
\author{Tian-Zi Song}
\affiliation{School of Physics, Sun Yat-Sen University, Guangzhou 510275, China}
\author{Xue-Sen Wang}
\author{Yu-Mei Zhang}
\email[Corresponding author, ]{zhangym26@mail.sysu.edu.cn}
\affiliation{Sino-French Institute of Nuclear Engineering and Technology, Sun Yat-Sen University, Zhuhai 519082, China}
\author{Tao Lin}
\email[Corresponding author, ]{lintao@ihep.ac.cn}
\affiliation{Institute of High Energy Physics, Chinese Academy of Sciences, Beijing 100049, China}
\author{Zheng-Yun You}
\email[Corresponding author, ]{youzhy5@mail.sysu.edu.cn}
\affiliation{School of Physics, Sun Yat-Sen University, Guangzhou 510275, China}

\begin{abstract}
    In high-energy physics~(HEP) experiments, visualization software plays a pivotal role in detector design, offline software development, and event data analysis. The visualization tools integrate detailed detector geometry with complex event data models, providing the researchers with invaluable insights into experimental results. Phoenix is an emerging general-purpose visualization platform for the current and next-generation HEP experiments. In this study, we develop an event display software based on Phoenix for the CEPC experiment. It offers necessary functionalities for visualizing detector geometries and displaying event data, allowing the researchers to optimize detector design, test simulation and reconstruction algorithms, and analyze event data in a visualized way. Additionally, we discuss the future applications of the event display software, including its usage in online monitoring and the potential to build virtual reality projects for enhanced data visualization. 
\end{abstract}


\maketitle

\section{Introduction}
The Circular Electron Positron Collider~(CEPC) is an international scientific facility proposed by the Chinese particle physics community in 2012~\cite{CEPC_collider}. Designed to explore fundamental physics, particularly as a Higgs factory, the CEPC will be situated in China within a circular underground tunnel approximately 100 kilometers in circumference. This double-ring collider will circulate electron and positron beams in opposite directions through separate beam pipes, with detectors installed at two interaction points.

The CEPC is designed to operate in three distinct modes: H mode~($e^+e^-\to ZH$), Z mode~($e^+e^-\to Z$), and W mode~($e^+e^-\to W^+W^-$). These modes correspond to center-of-mass energies of 240 GeV, 91 GeV, and 160 GeV, respectively~\cite{CEPC_conceptual_design1, CEPC_conceptual_design2}. It will be an advanced factory of Higgs, Z, W bosons and top quarks, enabling precision measurements of the Higgs boson properties~\cite{HiggsStudyAtCollider}, along with those of the mediators of the weak interaction, the W and Z bosons~\cite{ElectroWeakTest,CPVZFactory}. It also provides an opportunity to search for physics beyond the standard model~(BSM), allowing scientists to discover new physics and unknown phenomena~\cite{CEPCPhysical}.

As the CEPC project progresses~\cite{CEPC_status}, a series of software needs to be developed, and the event display software is an indispensable part~\cite{HEPSoftwareRoadmap}. In modern High-Energy Physics~(HEP) experiments, the structures of detectors are becoming increasingly complex to achieve higher measurement accuracy and detection efficiency. This fact necessitates sophisticated visualization tools, to aid scientists learn about the structure of detectors, analyze physical events, and optimize simulation and reconstruction algorithms~\cite{CEPCReconstruction, CEPCReconstructionInDC, TrackReconstruction}. 

To develop event visualization tools for HEP experiments, the ROOT software~\cite{ROOT_framework} and its EVE package~\cite{ROOT_EVE} have been widely used. However, as a native C++ application, it lacks advanced web-based features such as cross-platform compatibility and ease of deployment. The JSROOT~\cite{jsroot} package, which is a web-based extension of ROOT, addresses many of these limitations by enabling browser-side visualization through WebGL and JSON data formats. However, there still lacks a unified framework that integrates modern 3D rendering capabilities with user-friendly interfaces. Phoenix is a web-based event display framework that is specifically designed for HEP experiments. It is lightweight and highly cross-platform, built on top of JSROOT and three.js libraries~\cite{threejs}. With the aid of the two components, Phoenix can provide detailed 3D display of detector geometry and events along with a user-friendly interface, making it a good choice for developing visualization tools for CEPC. Therefore, in this study we develop an event display software for CEPC using the Phoenix framework. The software not only provides basic visualization functions for detector and events, but also has a variety of extra features, allowing the users to change the visualization effects through graphical user interface~(GUI) and check tracks, hits and Monte Carlo~(MC) information. Hence, the Phoenix based event display software will be an effective tool for detector design, offline software development, and physics analysis in the CEPC experiment.

The rest of this paper is structured as follows.
In Section~\ref{sec:Phoenix framework}, we introduce the Phoenix framework for visualization in HEP experiments. 
In Section~\ref{sec:Methodologies}, the structure and data flow of event display for CEPC based on Phoenix is presented.
In Section~\ref{sec:Visualization}, the visualization features of the CEPC event display are described.
The applications are discussed in Section~\ref{sec:Features and applications}.
Finally, Section~\ref{sec:Summary} gives the summary.   

\section{Phoenix framework}
\label{sec:Phoenix framework}

\begin{figure*}
    \centering
    \includegraphics[width=0.8\linewidth]{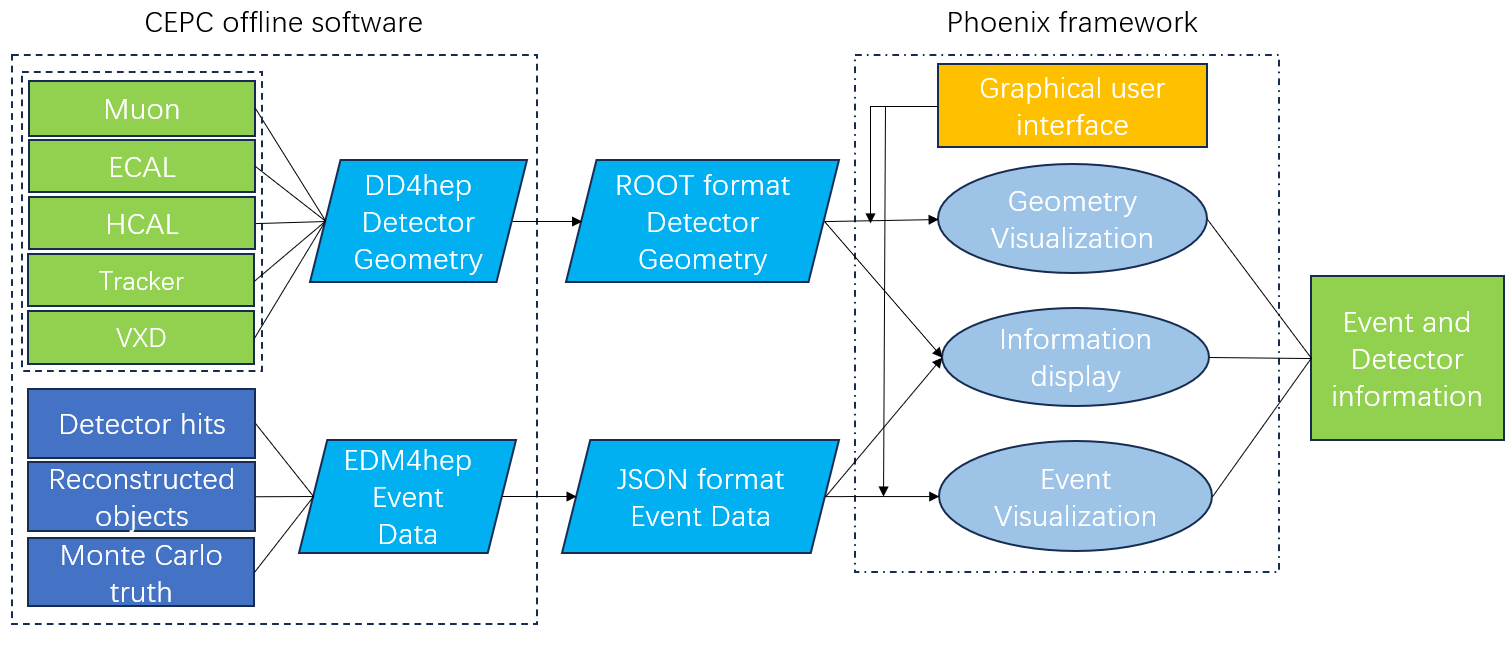}
    \caption{The structure and data flows of CEPC event display software based on Phoenix. The left shows the CEPC offline software, which provides detector geometry in DD4hep compact XML format and event data in EDM4hep format. The geometry and event data are converted to ROOT format and JSON format, respectively, serving as the input of the Phoenix framework on the right. With a graphical user interface, Phoenix provides a detailed 3D visualization and information display.}
    \label{fig:display_structure}
\end{figure*}

Phoenix is a web-based event visualization framework for HEP experiments~\cite{phoenix}. It is based on the TypeScript language and Angular.js framework~\cite{angularjs}, and uses the popular three.js~\cite{threejs} library for 3D support. In 2017, the HEP Software Foundation~(HSF) visualization white paper called for a common event format and common tools to aid visualization~\cite{HSF_whitepaper}, and the Phoenix framework was adopted to be an extensive framework for event and geometry display. The Phoenix framework currently supports built-in geometries and event data for ATLAS~\cite{ATLAS}, CMS~\cite{CMS}, LHCb~\cite{LHCb}, and TrackML~\cite{TrackML}. Several demonstrations have also been created for FASER~\cite{FASER}, FCC-ee, and FCC-hh~\cite{FCC}.

Compared to native applications built on ROOT software and its EVE package, the Phoenix framework has the following advantages.
\begin{itemize}
    \item \emph{Highly cross-platform} Unlike other event display software built as native C++ applications, Phoenix is a web-based framework and can run on any modern web browser. This framework is built as a web application and can be opened by simply visiting a URL.

    \item \emph{High-level support for HEP experiments} As an event visualization framework for HEP experiments, Phoenix provides high-level support for visualizing events. It supports visualizing various types of physics objects, including tracks, hits, calorimeter cells, vertices, compound objects and missing energy. 
    Phoenix also has integrated the JSROOT~\cite{jsroot} package, which is part of the ROOT framework, to provide support for visualizing 3D ROOT geometry. In addition to the ROOT format, displaying 3D geometries like OBJ~\cite{OBJformat}, glTF~\cite{gltf}, and JSON are also supported.

    \item \emph{Extensive} Designed as experiment agnostic, Phoenix aims to support as many experiments as possible. It provides a standard interface to create a visualization application, thus adding a new experiment demonstration only requires the developer to create a simple extension.

    \item \emph{User interface and features} Phoenix offers rich user interface options, including labeling functionality for physics objects, shareable URLs for predefined events and configurations, useful outreach activities or physics briefings, and built-in animations that make event displays interactive and engaging. All these configurations can be saved, reloaded, applied by default, or passed through URL parameters.
\end{itemize}

Therefore, the Phoenix framework is a competitive choice for developing event visualization tools. In fact, Phoenix has been the official web event display for ATLAS, and is also used by FCC, LHCb, and Belle II experiments~\cite{LHCb_Phoenix}.

\section{Methodologies}
\label{sec:Methodologies}

An event display software for HEP experiments should provide detailed visualization of detector structure and event data, featuring to show important information, such as detector hits, showers, reconstructed tracks, event information, and a GUI to control the display. To achieve this, the CEPC event display software adopts the Phoenix framework and uses the CEPC offline software~\cite{CEPCSW} to provide the detector geometry and event data. 

The structure of the Phoenix-based CEPC event display software is shown in Fig.~\ref{fig:display_structure}.
The workflow of CEPC event display software can be divided into the following steps. Firstly, the CEPC offline software provides detector geometries in DD4hep~\cite{DD4hep, DD4hep2} compact XML format. It also generates event data through simulation or reconstruction and stores them in EDM4hep~\cite{EDM4hep, EDM4hep_2} format. DD4hep and EDM4hep are both parts of Key4hep, a framework providing all the necessary ingredients for future HEP experiments~\cite{Key4hep,STCFSoftware}. Secondly, since the Phoenix framework does not directly support DD4hep XML format geometries and EDM4hep format event data, a series of conversions are needed~\cite{GeometryConversionToUnity, DD4hepConversion}. The geometries are converted to ROOT format using the ROOT software, and the event data is converted to JSON format while preserving essential information for visualization. Finally, the geometry files and event data are loaded to the web display tool, rendering the detector and events, while essential information for analysis like particle energy or tracker hits are also displayed in the user interface~(UI). 

\begin{figure}
    \centering
    \includegraphics[width=1\linewidth]{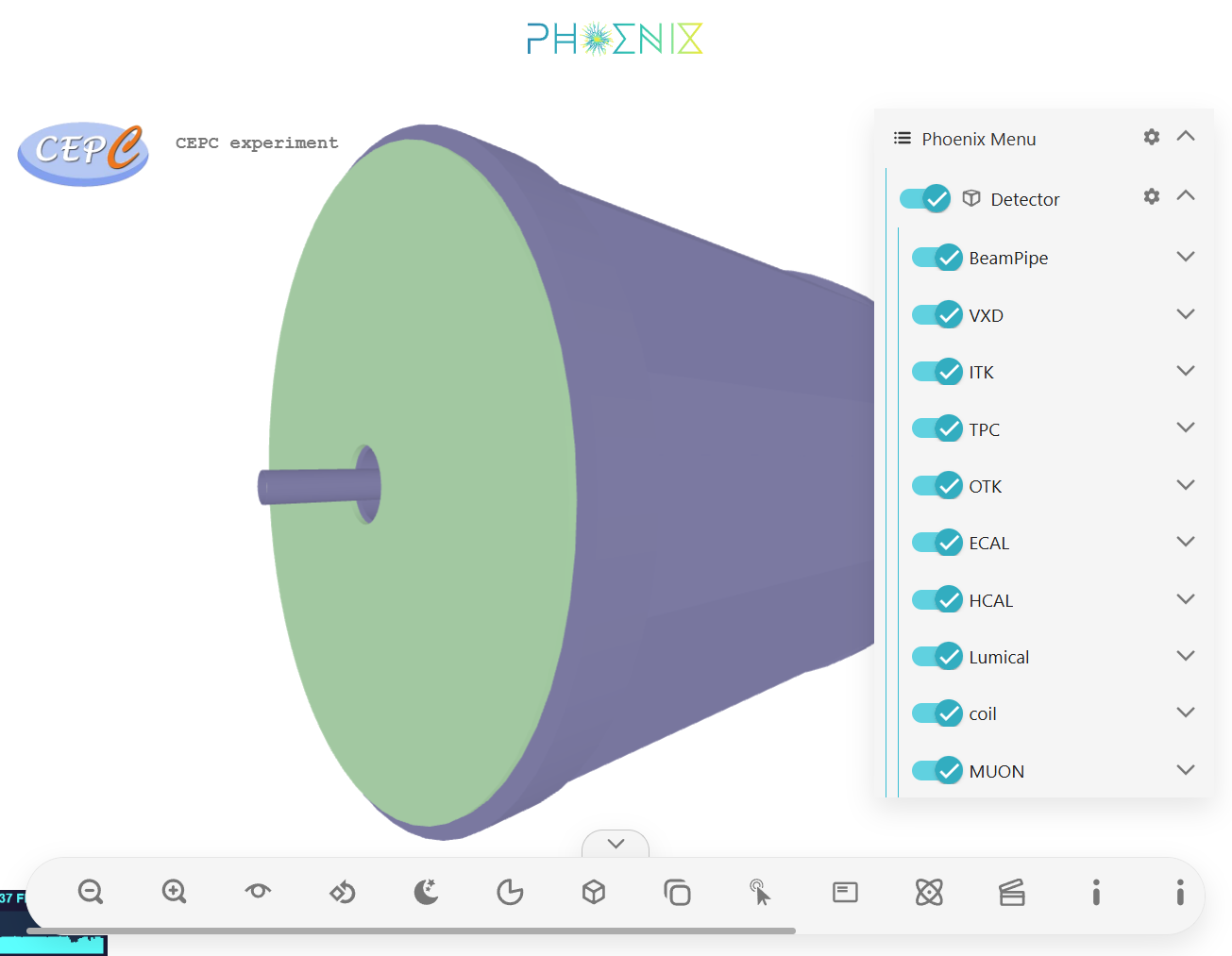}
    \caption{The GUI of CEPC event display software based on Phoenix. The detector is shown in the center with the UI components in the surrounding area. The panel on the right is the Phoenix menu, and the tool bar on the bottom is the Phoenix icon-bar.}
    \label{fig:phoenix_GUI}
\end{figure}

A simple but practical GUI is available for the user to control the workflow. Fig.~\ref{fig:phoenix_GUI} shows the user interface of the event display software. The right corner of figure is the Phoenix menu, which can be used to control the visibility of each detector object; the bottom of figure shows the Phoenix icon-bar, which can be used to upload or export files, change display modes, enable or disable other high-level features. Users can also use the GUI control to upload geometry and event data files for visualization, change the display effects, and choose which part of the event or detector to be visualized or not. The CEPC and Phoenix logos are displayed on the left and the top, and users can visit the CEPC and Phoenix homepage via clicking the logos.

\section{Visualization}
\label{sec:Visualization}

\subsection{Visualization of detector}
To enable visualization of the CEPC detector within the Phoenix framework, a series of conversion steps are undertaken~\cite{GEANT4_Conversion}. Initially, the geometries of the detector are stored in DD4hep XML description within the CEPC offline software. By running Geant4~\cite{GEANT4} simulations, the detector geometries are loaded to perform simulations, and with the G4GDMLParser module provided by the Geant4 toolkit, the geometries can be exported from the simulation in Geometry Description Markup Language~(GDML) format~\cite{GDML, BesGDML}. This step ensures that all geometric details of the detector components are accurately captured and preserved. 

Subsequently, these GDML geometries are converted into ROOT format using the geometry export functionalities provided by ROOT framework. This conversion is necessary because Phoenix does not natively support GDML as a detector input format. The conversion process from DD4hep XML format to ROOT format retains the full geometric complexity of the detector, with no explicit simplification applied to the original design. Although the ROOT format internally represents geometry using Constructive Solid Geometry (CSG), the JSROOT package automatically converts this CSG-based geometry into surface-based representations for rendering. To manage performance during 3D visualization, JSROOT also allows setting upper limits on the number of visible nodes and surfaces, effectively mitigating potential rendering bottlenecks. As a result, even highly complex detector models can be visualized interactively without significant performance degradation.

In the context of the CEPC detector visualization within the event display software, the detector is composed of several key components, including but not limited to the Vertex Detector~(VXD)~\cite{CEPC_VXD}, the Tracker~\cite{CEPC_TPC, CEPC_DC}, the Electromagnetic Calorimeter~(ECAL)~\cite{CEPC_ECAL}, the Hadron Calorimeter~(HCAL)~\cite{CEPC_HCAL} and the Muon Detector~(Muon). The geometry of each component is individually generated and loaded separately into Phoenix, allowing users to selectively visualize specific parts of the detector through the Phoenix menu.

\begin{figure}
    \centering
    \subfigure[Detector geometry in CDR]{\includegraphics[width=0.6\linewidth]{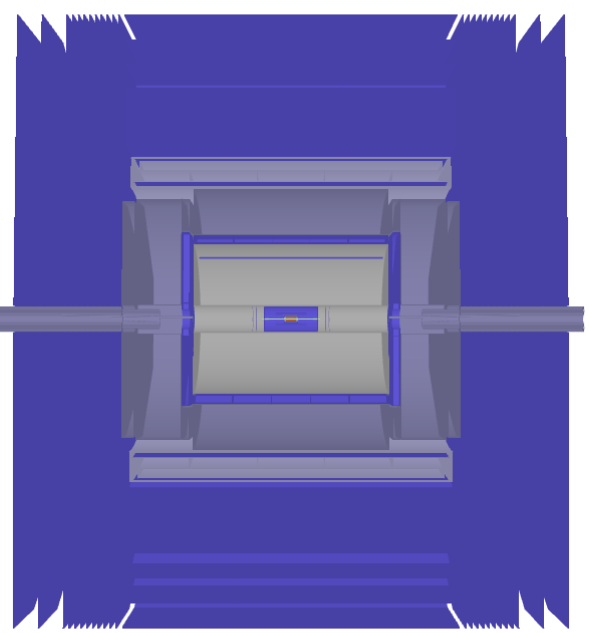}}
    \subfigure[Detector geometry in TDR]{\includegraphics[width=0.8\linewidth]{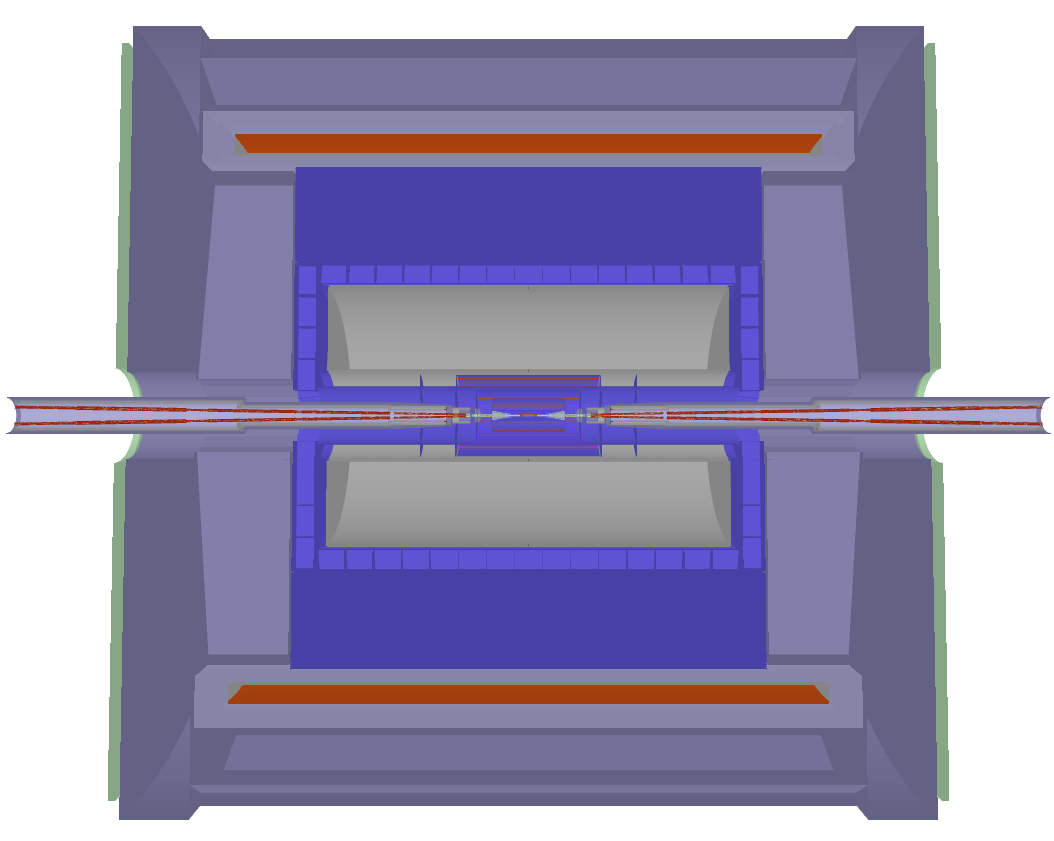}}
    \caption{The section views of different versions of CEPC detector geometries. (a) shows the Conceptual Design Report~(CDR) version of detector geometry, while (b) presents the latest Technical Design Report~(TDR) version of detector geometry.}
    \label{fig:CEPC_detector_Visualization}
\end{figure}

With the geometries loaded to Phoenix framework, they are immediately visualized in the software, and the users can freely change their perspective of view, zoom in or zoom out the 3D model, and clip the geometries to observe the details inside the detector. As the CEPC detector is still under design, multiple versions of geometry are available in the software for comparative analysis. Fig.~\ref{fig:CEPC_detector_Visualization} illustrates the visualization of several versions of the CEPC detector within Phoenix. Fig.~\ref{fig:CEPC_detector_Visualization}~(a) shows the conceptual design report~(CDR) version of detector geometry. Fig.~\ref{fig:CEPC_detector_Visualization}~(b) displays the latest technical design report~(TDR) version of detector geometry under development~\cite{CEPC_technical_design}. In addition, Fig.~\ref{fig:CEPC_detector_Visualization} presents detailed section views of CEPC detector geometries, sequentially showing the VXD, Tracker, ECAL, HCAL, Electromagnetic Coil~\cite{CEPC_coil}, and Muon detector from the innermost to the outermost layers~\cite{CEPC_conceptual_design2}.

\begin{figure}
    \centering
    \subfigure[Muon detector]{\includegraphics[width=0.43\linewidth]{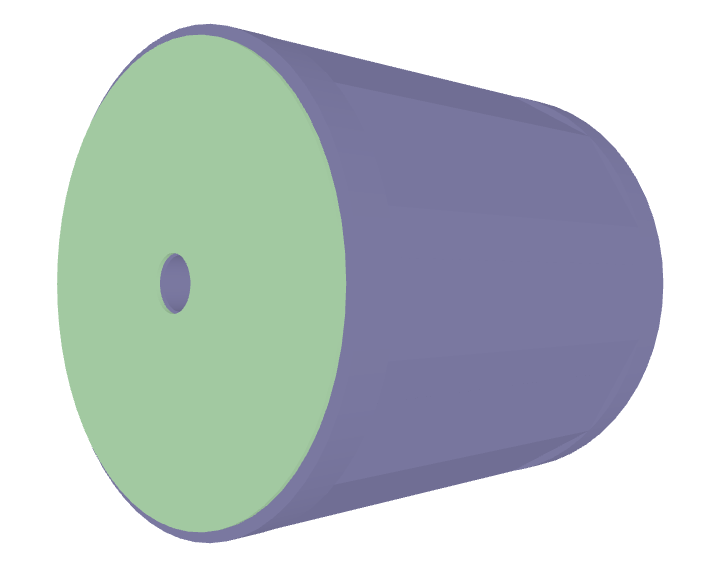}}
    \subfigure[Electromagnetic coil]{\includegraphics[width=0.43\linewidth]{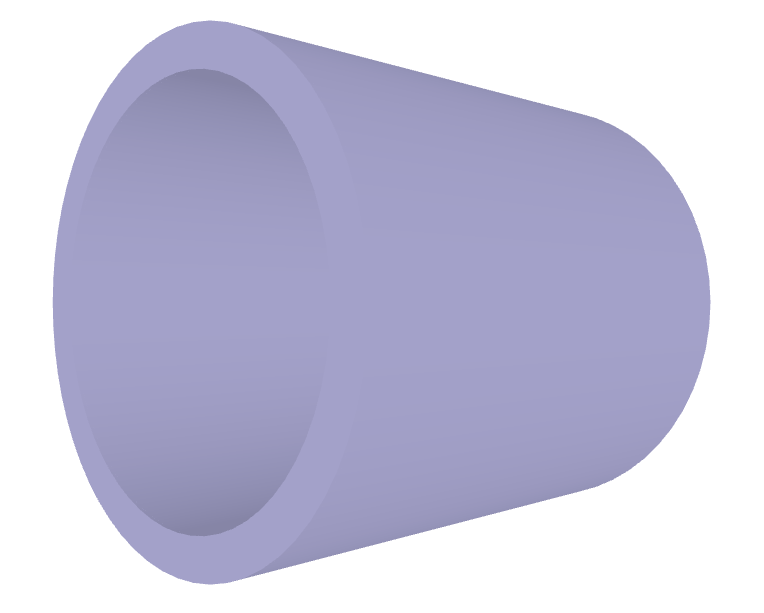}}
    \subfigure[ECAL]{\includegraphics[width=0.43\linewidth]{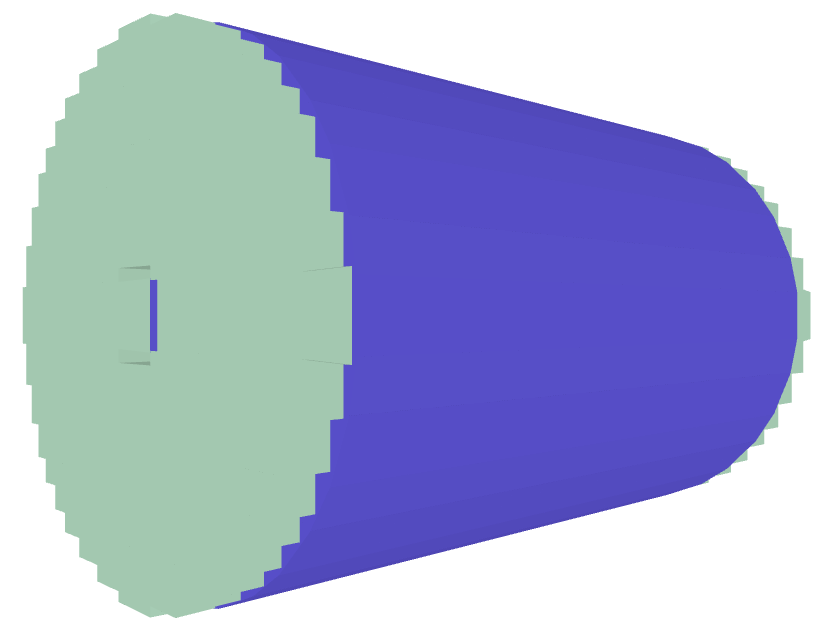}}
    \subfigure[HCAL]{\includegraphics[width=0.43\linewidth]{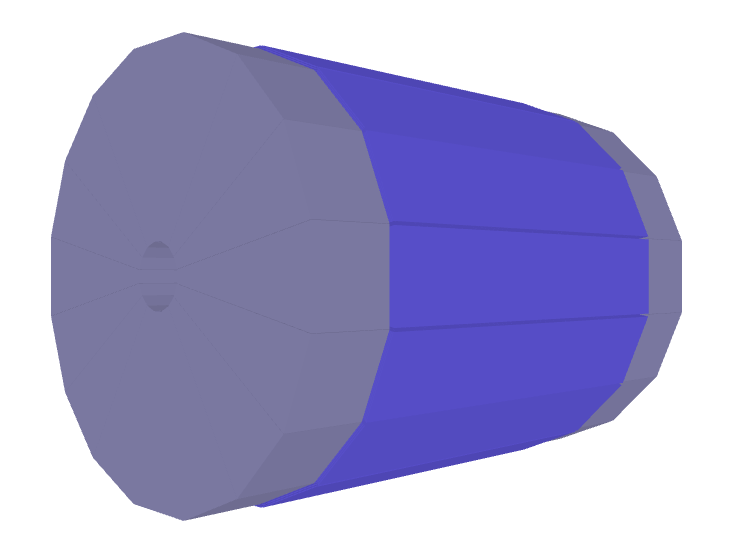}}
    \begin{minipage}{1.0\linewidth}
    	\centering
    	\includegraphics[width=0.7\linewidth]{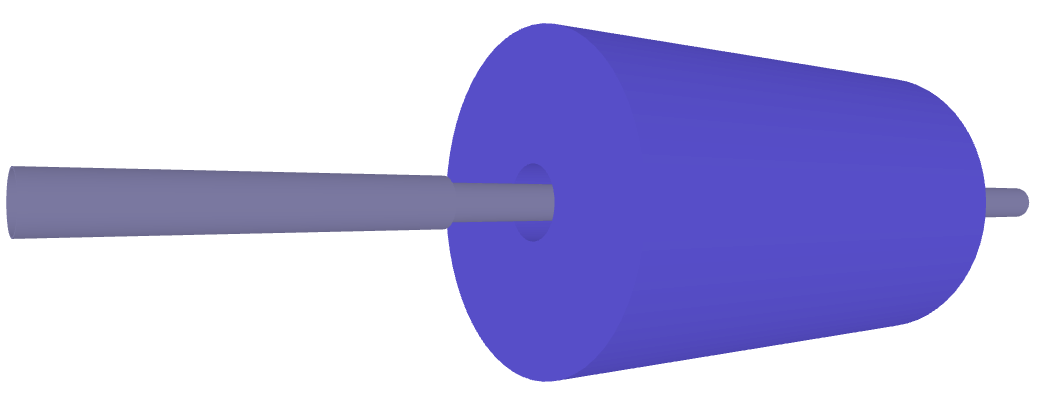}\\
    	(e) Tracker
    \end{minipage}
    \caption{Visualization of various sub-detectors in CEPC event display software.}
    \label{fig:CEPC_subdetectors}
\end{figure}

\begin{figure}
    \centering
    \includegraphics[width=0.7\linewidth]{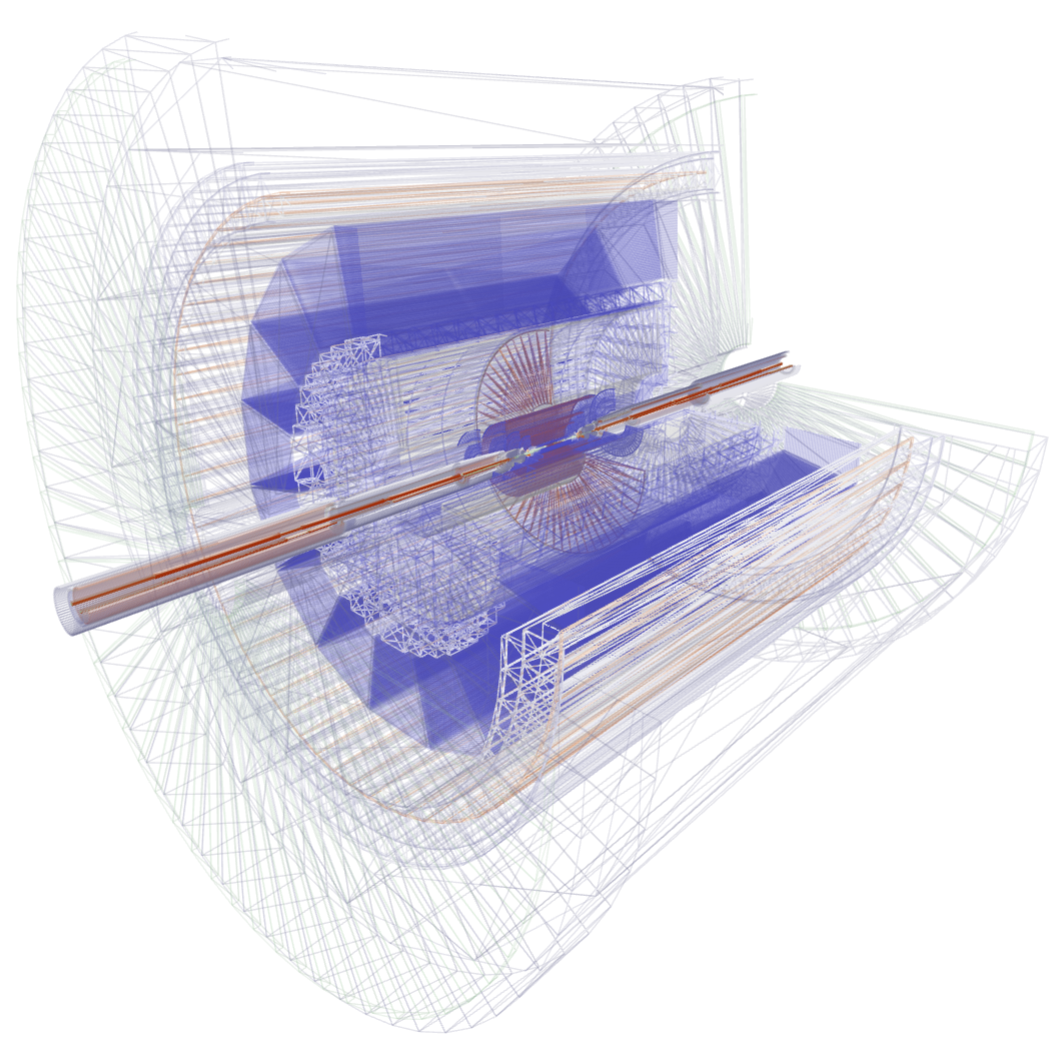}
    \caption{Visualization of CEPC detector in the wire frame mode.}
    \label{fig:CEPC_Wireframe_Visualization}
\end{figure}

Additionally, the Phoenix framework offers users the flexibility to select specific sections of the detector for visualization. For instance, Fig.~\ref{fig:CEPC_subdetectors} illustrates the display of CEPC sub-detectors. Due to performance constraints, many detailed aspects of the detector geometries may be omitted in the default view. However, users can manually change the opacity of 3D model, or activate the "Wireframe" mode through the user interface, as shown in Fig.~\ref{fig:CEPC_Wireframe_Visualization}. The mode reveals the mesh structures of the detector geometries, thereby providing a detailed view of the geometric configurations while also rendering the 3D models transparent. This feature is particularly useful for gaining deeper insights into the internal structure and layout of complex detector components.

\subsection{Visualization of events}
In HEP experiments, it is standard practice to store event data in ROOT format due to its efficiency and flexibility. However, since Phoenix does not natively support ROOT event format for visualization, an additional step involving the conversion of event data formats is required before visualization can occur. The Phoenix framework provides support for loading event data from JSON format file. To convert ROOT event format to JSON format, the tool EDM4hep2json, which is provided by the EDM4hep~\cite{EDM4hep2json} toolkit, is utilized. This tool is capable of converting EDM4hep event data to JSON format while preserving the essential event information required for visualization.

Specifically, the EDM4hep2json toolkit effectively "dumps" the EDM4hep event data into JSON format, ensuring that all relevant event information is retained and usable within Phoenix. This conversion process bridges the gap between the event data storage format and the visualization requirements of the Phoenix framework, enabling comprehensive and interactive visualizations of the events in CEPC detector.

\begin{figure}
    \centering
    \subfigure[]{\includegraphics[width=0.49\linewidth]{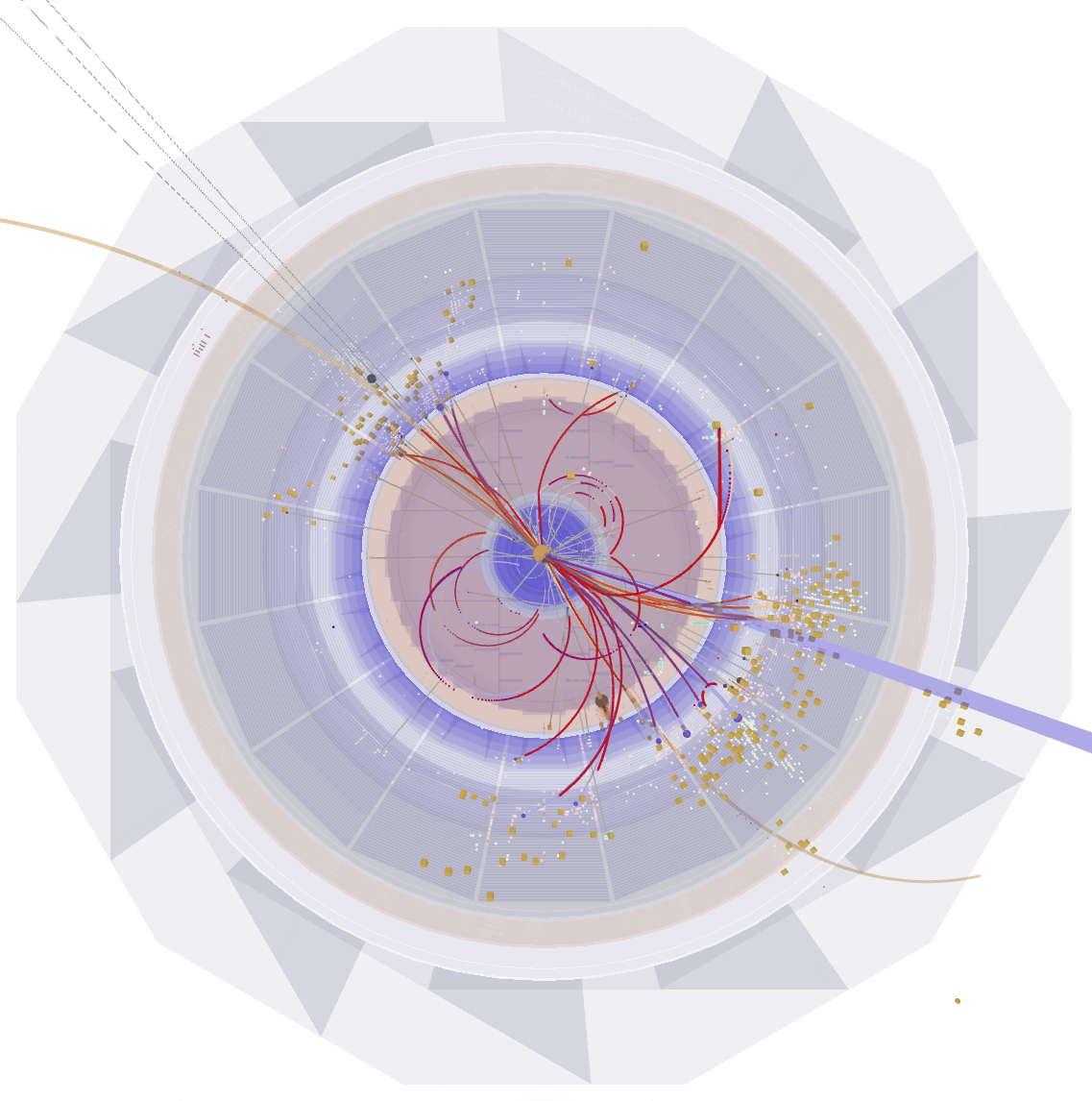}}
    \subfigure[]{\includegraphics[width=0.49\linewidth]{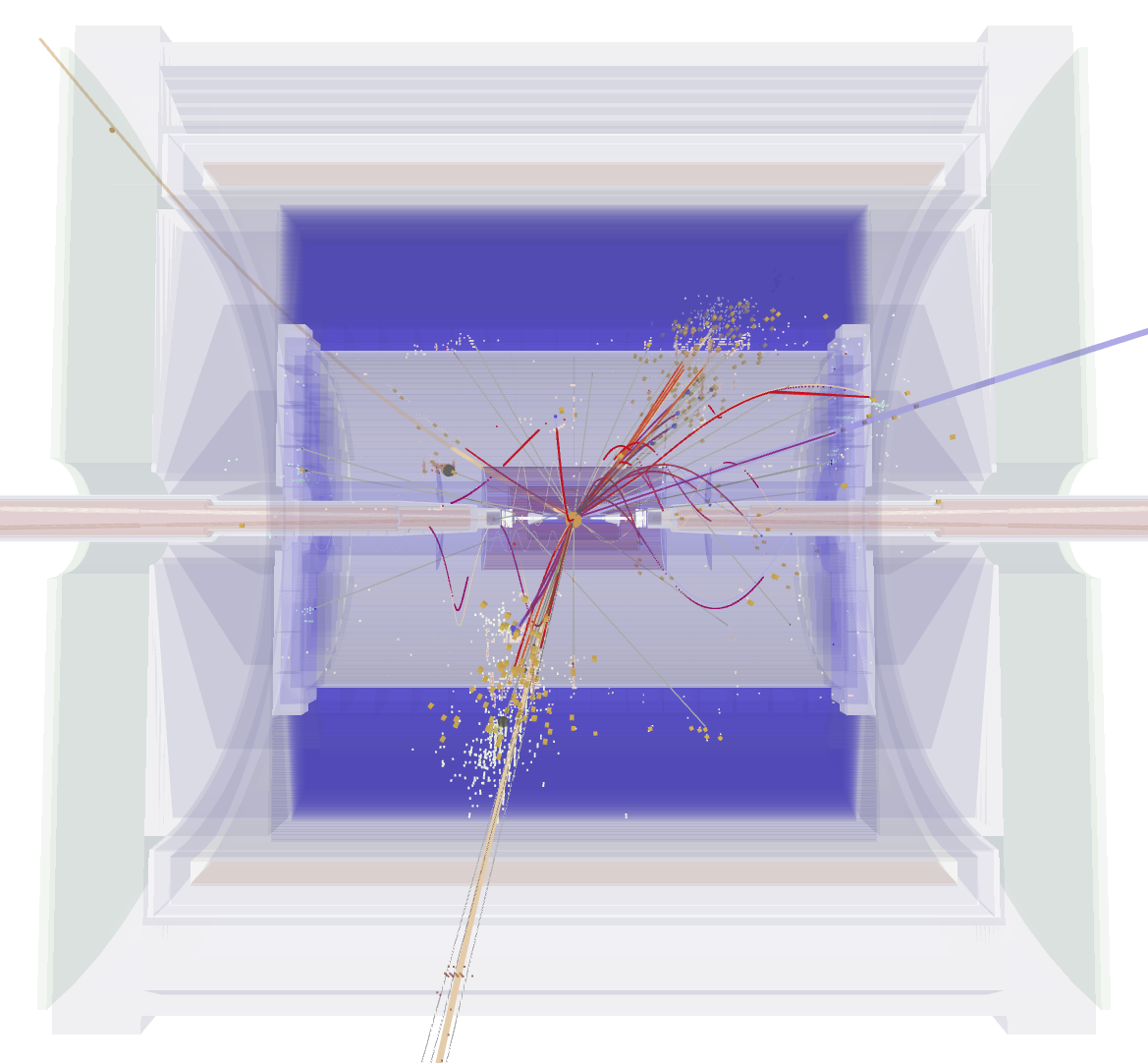}}
    \subfigure[]{\includegraphics[width=0.95\linewidth]{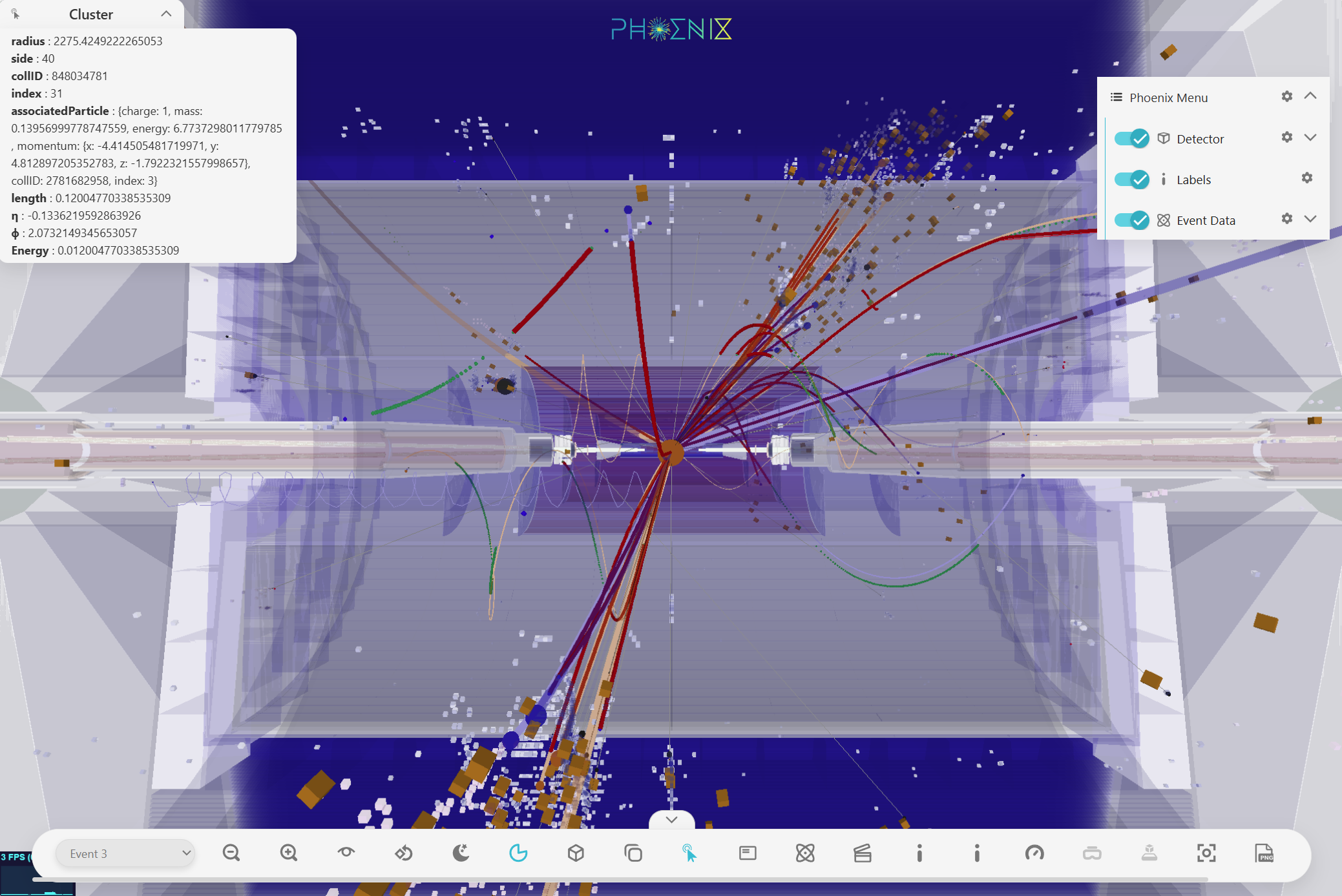}}
    \caption{The display of an $e^+e^-\to ZH, Z\to \mu^+\mu^-, H\to b\overline{b}$ event in CEPC event display. The detector is shown here to illustrate the interaction between particles and detectors. (a) and (b) plot the detector and event in X-Y and Z-R section view, respectively. (c) enlarges the details of (b) to show the event objects, where the red lines are tracks, the small and white cubes are calorimeter cells, and the yellow pointing cuboids are clusters.}
    \label{fig:CEPC_Event_Visualization}
\end{figure}

Once converted to JSON format, the user can manually upload the event file to the event display software, and the first event in the event data file will be immediately visualized. The user can also choose which event to visualize through the user interface. Fig.~\ref{fig:CEPC_Event_Visualization} provides an illustrative example of the CEPC detector along with a single event as visualized by the Phoenix event display software. The figure highlights three primary physical event objects:

\begin{itemize}
\item \emph{Tracks} Representing the trajectories of charged particles, depicted as red lines.
\item \emph{Calorimeter cells} Indicating energy deposits within a calorimeter, shown as small and transparent cubes.
\item \emph{Clusters} Groupings of energy deposits in a calorimeter, represented by yellow pointing cuboids.
\end{itemize}

In addition to these, the software supports the visualization of several other physics objects, including:

\begin{itemize}
\item \emph{Jets} Simplified geometric cones indicating the direction of energy flow.
\item \emph{Hits} Individual detector measurements, illustrated as points in space.
\item \emph{Vertices} Origins of tracks, displayed as dots.
\end{itemize}

The software further enhances user experience by enabling selective visualization of event objects through its intuitive interface. Users can disable certain object types, such as calorimeter cells, to focus on higher-level features like clusters. For instance, if the clusters within the ECAL detector are stored as a collection in the EDM4hep event data file, users have the option to enable or disable their visualization, allowing for focused analysis on specific aspects of the event.

Moreover, the software facilitates detailed analysis by providing information about selected objects. When "Object selecting" mode is activated, users can select objects to view detailed information in the info panel located on the top left corner. For example, selecting a cluster reveals its position, energy, and associated reconstructed particle properties~(including charge, energy, mass, and momentum), as shown in the left-top corner of Fig.~\ref{fig:CEPC_Event_Visualization}~(c). This functionality is invaluable for thorough event analysis.

\subsection{Visualization of simulation information}
The Monte Carlo information, which is generated from simulated particle interactions, provides ground-truth data for validating reconstructed objects (e.g., matching tracks to MC particles). In modern HEP experiments, MC information also plays an important role in event analysis, detector design and algorithm optimizing~\cite{CEPCReconstructionInDC, DCSimulation}. It can help physicists evaluate the performance of detector, test the reconstruction and analysis processes, understand the signal event and source of backgrounds. Since the CEPC experiment is still under planning, MC simulation is the primary source of event data, and utilizing MC information is the key of optimizing detector design and reconstruction algorithm~\cite{TrackReconstruction}.

\begin{figure}
    \centering
    \subfigure[Reconstructed information]{\includegraphics[width=0.7\linewidth]{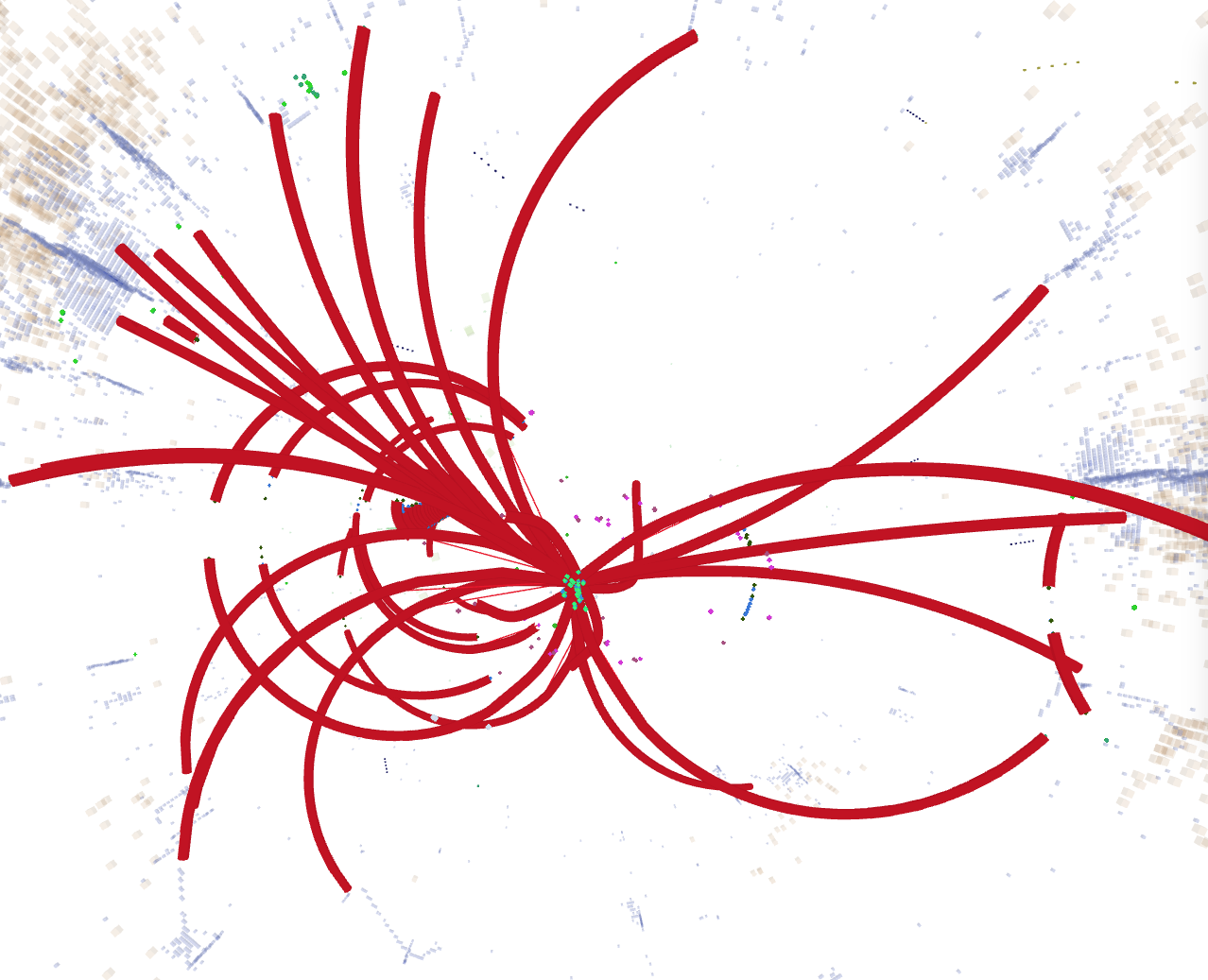}}
    \subfigure[MC truth information of particles]{\includegraphics[width=0.7\linewidth]{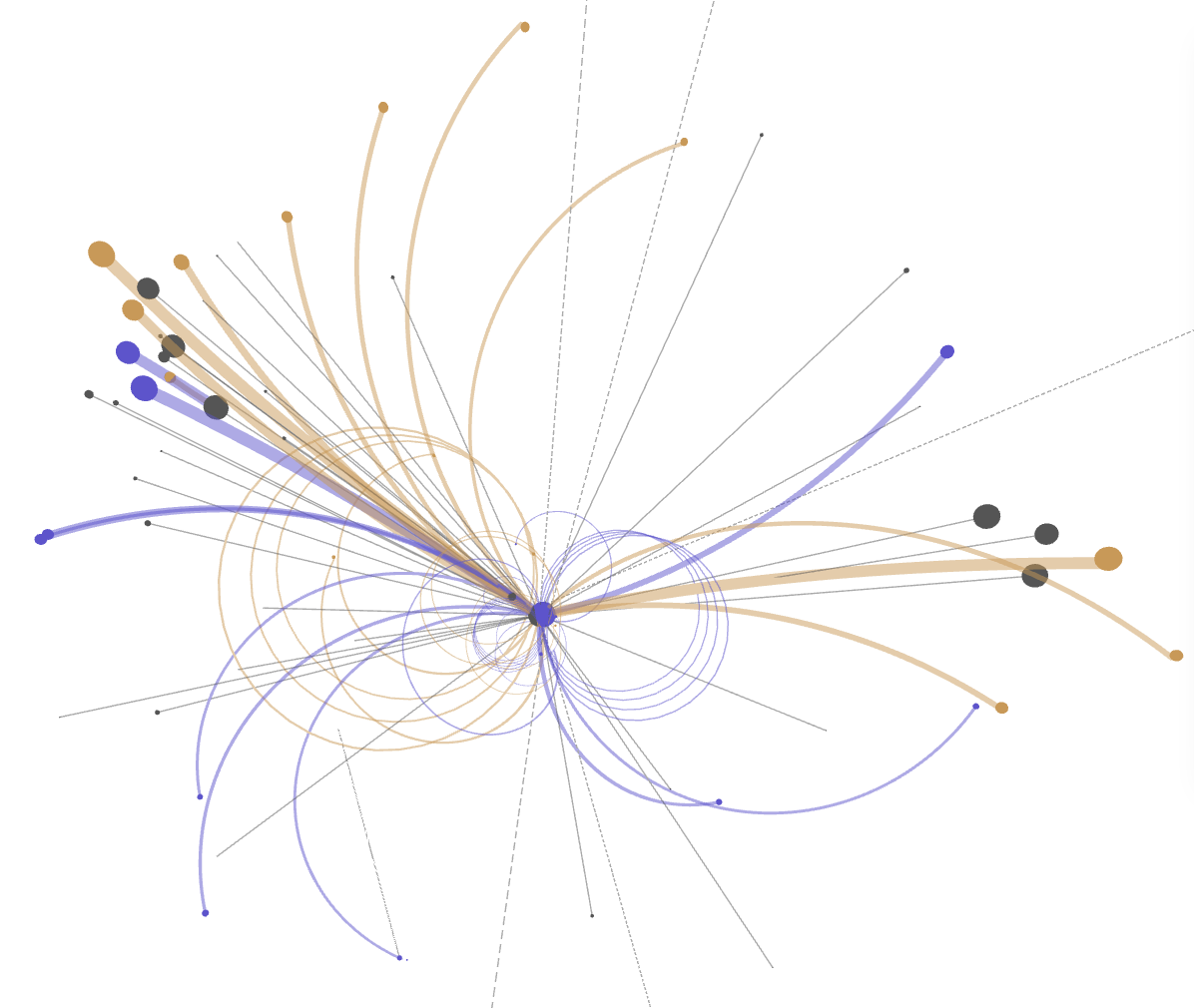}}
    \caption{Display of the reconstructed information and MC truth of a simulated event $e^+e^-\to ZH, Z\to v\overline{v}, H\to gg$, (a) plots the reconstructed tracks, calorimeter cells and hits in the event, while (b) displays the MC truth information of all particles in this event.}
    \label{fig:CEPC_MC_information_display}
\end{figure}

On the basis of Phoenix framework, we extend the original set of visualization primitives to display particles. Now our event display software supports detailed visualization of MC information. The EDM4hep event data model stores MC information as an individual collection, and the software will display it along with other event objects. For collider experiments such as CEPC, the MC information includes detailed properties of MC simulated particles, enabling physicists to validate reconstruction processes by comparing reconstructed tracks with MC truth information of particles. The event display software extracts information about MC particles, such as their vertex, momentum, charge, and endpoint, to calculate and draw their trajectories within the detector. As an example, Fig.~\ref{fig:CEPC_MC_information_display} displays an event with its reconstructed information and particle MC truth information together. The similarity of the two sub-figures distinctively validates the correctness of reconstruction from visual comparison.

In the event display software, the MC truth information is visualized as a set of tracks. To obtain a better visualization effect, the software divides MC particles into three categories.

\begin{itemize}
    \item \emph{Charged particles} The charged particles will interact with the magnetic field and be detected by the Tracker. They are drawn as blue and yellow curves to show their possible track according to MC information. The curves are interpolated to show their trajectories more precisely in the magnetic field. The colors of curves represent the charge, negatively charged particles are drawn as yellow curves, while positively charged particles are assigned with blue color.

    \item \emph{Neutral particles~(except for neutrinos)} The neutral particles may interact with calorimeters, leaving energy deposits in the detector. They are plotted as straight black lines.

    \item \emph{Neutrinos} In most cases, neutrinos will not interact with the detector, and they are always seen as the missing energy and missing momentum in event analysis. The software uses black dashed lines to represent neutrinos.
\end{itemize}

Also, the tracks visualize the complete spatial information of MC particles, including their vertices and endpoints. The endpoint of each track is marked by a sphere, which can help physicists to determine the moving direction of MC particles. Furthermore, like the other event objects, the display of MC information can be enabled or disabled through user interface, and the tracks of MC particles can also be selected to provide more detailed information.

\section{Features and applications}
\label{sec:Features and applications}

\subsection{Features compared to other event display software}
The CEPC event display software is forked from the original Phoenix framework and is developed as an independent branch, which not only inherits the advantages and features of Phoenix framework, but also introduces some new functionalities to meet the requirements for detector design and event analysis. Compared to the original Phoenix framework, the CEPC event display software provides full support for visualizing detector geometries and events, it also has some extra features to provide more diverse visualization and information, such as the display of MC truth information and enhanced geometry clipping capabilities. Compared to the ROOT based~\cite{BesVis, JunoROOT, JunoTaoROOT} or Unity based~\cite{Unity, JunoUnity, BesUnity} event display software, its advantages become even more obvious, making it the prime choice for CEPC event display in some specific situation. Generally speaking, the CEPC event display software has the following prominent features.

\emph{Easily accessible} Based on three.js, the event display software can be either deployed locally or on a web server, and then the user can access it with any web browser that supports WebGL~\cite{WebGL}. This includes most of the modern web browsers, such as Chrome, Firefox, Edge, Android browser and iOS Safari. Similar to other software, browsers rely on GPUs to accelerate rendering 3D models, and thus the event display software requires modern GPU support. However, with the aid of advancing front-end technologies, it minimizes the need for powerful hardware, and can operate efficiently even on mobile devices.

\emph{Lightweight} One of the outstanding features of the Phoenix based CEPC event display software is its lightweight nature. Unlike the ROOT framework, which heavily relies on third party libraries, it can be easily bundled into a package and distributed to other devices. The software bundle occupies only 85~MB of storage space, enabling the users deploy the application on a personal computer without installing external software packages. Meanwhile, the Unity editor also allows to build the source code into a standalone application. However, the resulting applications are generally not cross platform and usually occupy too much memory.

\emph{Fast} The CEPC event display software also demonstrates great advantages on performance. To evaluate its performance, we deploy the software on a personal computer with an Intel i5-12400 processor and an Intel Iris Xe integrated graphics card, and access it via the Chrome browser. The software renders the detector and event smoothly, achieving a refresh rate exceeding 30 frames per second~(FPS), and the average memory usage is about 800~MB. During the test, we observe that the Iris Xe GPU utilization increased from 1\% to approximately 18\% during active visualization, indicating that the rendering process leverages both CPU and GPU resources for efficient performance. Notably, the performance is closely related to the complexity of detector geometry and event data. When rendering simpler geometries, the refresh rate can be even higher. In comparison, the typical refresh rate of ROOT-based event display software is about 10~FPS~\cite{JunoROOT, JunoTaoROOT}, and a high latency between the users and computing servers can significantly reduce its performance.

\emph{Comprehensive visualization features} Instead of writing basic logics for event display, the CEPC event display software inherits the functionalities of Phoenix framework, which provides complete features for detector and event visualization. Furthermore, the software extends the original set of functions of Phoenix framework to fulfill the needs for deeper analysis, making it a practical choice for CEPC event display.

\subsection{Potential applications}
As previously introduced, the CEPC event display software provides a convenient and flexible way to visualize detector and events. It has great potential for applications and further developments, some possible applications of the event display software are listed below.

\emph{Detector design and event analysis} As an event display software, the primary goal of its development is to assist researchers to analyze detector geometry and event data. The event display software provides necessary features to display CEPC events and detectors, and it also has the flexibility to display other 3D geometries with EDM4hep format events, enabling the users to gain insights about detector structure and event signal. The researchers can receive feedback from the software and consequently optimize detectors and algorithms.

\emph{Online monitoring} Since CEPC event display software is developed using the web-based Phoenix framework, it is naturally suitable for online tasks. Instead of writing complex logic to fetch data from the internet, the event display software can be easily extended to visit URLs to display online event data. Since the software can run in any modern web browser, the users can even check the experiment status with a smart phone. This will definitely help the physicists to instantly modify the status of facility and enhance high-quality experimental data taking.

\emph{Further development} As introduced before, the Phoenix framework is highly modular and extendable, the changes and extensions we develop for the event display software can be easily applied to other applications, even to other HEP experiments such as BESIII~\cite{BESIII} and JUNO~\cite{JUNO}. Moreover, the 3D library three.js~\cite{threejs} used by Phoenix has provided support for Virtual Reality~(VR)~\cite{VirtualReality} and Augmented Reality~(AR)~\cite{AugmentedReality} features, one can further develop the software to build VR/AR applications for the CEPC experiment.

\section{Summary}
\label{sec:Summary}

With the aid of Phoenix framework, we develop a web-based event visualization software for the CEPC experiment. It is purely based on front-end web technologies, and can be easily deployed and run on any web server or personal computers. This software provides necessary facilities for visualizing and displaying CEPC detector and events, allowing researchers to gain deep insights into detector structures and event characteristics, enhancing continuous optimization of detectors, simulation and reconstruction algorithms. It also has great potentials for a variety of applications, such as online monitoring and VR/AR programs. 

\section*{Acknowledgements}
This work was supported by the National Natural Science Foundation of China (Nos. 12175321, W2443004, 11975021, 11675275, and U1932101), National Key Research and Development Program of China (Nos. 2023YFA1606000 and 2020YFA0406400), National College Students Science and Technology Innovation Project, and Undergraduate Base Scientific Research Project of Sun Yat-sen University.

\printbibheading[heading=bibintoc]
\printbibliography[heading=none]

\end{document}